\documentclass[10pt,english,aps,manuscript]{article}
\usepackage[LGR,LGR,LGR,LGR,T1]{fontenc}
\usepackage[latin9]{inputenc}
\usepackage{textcomp}
\usepackage{amsmath}
\usepackage{amsthm}
\usepackage{graphicx}

\makeatletter

\DeclareRobustCommand{\greektext}{%
  \fontencoding{LGR}\selectfont\def\encodingdefault{LGR}}
\DeclareRobustCommand{\textgreek}[1]{\leavevmode{\greektext #1}}
\ProvideTextCommand{\~}{LGR}[1]{\char126#1}

\providecommand{\tabularnewline}{\\}

\numberwithin{equation}{section}
\numberwithin{figure}{section}
\newcommand{\lyxaddress}[1]{
\par {\raggedright #1
\vspace{1.4em}
\noindent\par}
}

\usepackage{babel}

\usepackage{babel}

\makeatother

\usepackage{babel}
\begin{document}

\title{Predictions for the isolated diphoton production through NNLO in
QCD and comparison to \linebreak{}
the 8 TeV ATLAS data }

\author{Bouzid BOUSSAHA$^{(\mathrm{a})}$\thanks{bo.boussaha@gmail.com},
Farida IDDIR$^{(\mathrm{a})}$\thanks{iddir.farida@univ-oran.dz}
and Lahouari SEMLALA$^{(\mathrm{a,b})}$\thanks{semlala.lhaouari@esgee-oran.dz, l\_semlala@yahoo.fr}}
\maketitle

\lyxaddress{$^{(\mathrm{a})}$Laboratoire de Physique Théorique d'Oran (LPTO),
University of Oran1-Ahmed Ben Bella, Algeria; }

\lyxaddress{$^{(\mathrm{b})}$École Supérieure en Génie Électrique et Énergétique
d'Oran, Algeria. }

{\footnotesize{}Published in }\textit{\footnotesize{}Advances in High
Energy Physics }{\footnotesize{}journal: doi.org/10.1155/2018/4174602}\texttt{\footnotesize{}.}{\footnotesize\par}
\begin{abstract}
We present cross section predictions for the isolated diphoton production
in next-to-next-to-leading order (NNLO) QCD using the computational
framework MATRIX. Both the integrated and the differential fiducial
cross sections are calculated. We found that the arbitrary setup of
the isolation procedure introduces uncertainties with a size comparable
to the estimation of the theoretical uncertainties obtained with the
customary variation of the factorization and renormalization scales.
This fact is taken into account in the final result. 
\end{abstract}

\section{Introduction}

Considerable attention, both experimental and theoretical, has been
paid to the study of the diphoton productions. This process is relevant
for testing the Standard Model predictions and is of great importance
in Higgs studies. The diphoton final state is also important in new
physics researches: the extra-dimensions, the supersymmetry and the
new heavy resonances are three important topics among others.

The theoretical calculations are possible thanks to the codes DIPHOX
\cite{DIPHOX}, \textsc{ResBos} \cite{ResBos}, 2\textgreek{g}\textsc{Res}
\cite{2=00003D00003D0003B3Res}, 2\textgreek{g}NNLO~\cite{2=00003D00003D0003B3NNLO},
MCFM \cite{MCFM} and recently MATRIX \cite{MATRIX-main}.

In addition to the \textit{direct} production from the hard subprocess,
photons can also result from the fragmentation subprocesses of QCD
partons. The complete NLO one- and two-\textit{fragmentation} contributions
are implemented in DIPHOX. In \textsc{ResBos} only a simplified one-fragmentation
contribution is considered but the resummation of initial-state gluon
radiation to NNLL accuracy is included. Both DIPHOX and \textsc{ResBos}
implement the gg \textrightarrow \textgreek{gg} component, to LO
and NLO in QCD respectively. In the (NLO) MCFM calculations, the fragmentation
component is implemented to LO accuracy.

Thanks to the high rate of production of final diphoton pairs (considered
as relatively clean), experimentalists make precise measurements,
pushing the experimental uncertainties down to the percent level,
thus NLO calculations have become insufficient and therefore more
precise investigations are required in order to reproduce the data
and to provide a precise modeling of the SM backgrounds.

During the first run of the LHC (Run I), measurements of the production
cross section for two isolated photons at a center-of-mass energy
of $\sqrt{s}=7$ TeV is performed by ATLAS \cite{ATLAS7} and CMS
\cite{CMS7}, based on an integrated luminosity of 4.9 fb$^{-1}$
and 5.0 fb$^{-1}$ respectively. This is concluded by ATLAS \cite{ATLAS8}
at $\sqrt{s}=8$ TeV using an integrated luminosity of 20.2 fb$^{-1}$
which gives a much more accurate result.

In Ref.\cite{ATLAS8}, the authors reported that NLO calculations
fail to reproduce the data and even if there is improvement of the
result with 2\textgreek{g}NNLO, it remains insufficient.

Although the NNLO isolated diphoton production cross sections can
be calculated using the 2\textgreek{g}NNLO and MCFM public codes,
we used the most recent code MATRIX, because, in addition to its NNLO
accuracy, it allows us to estimate systematic errors related to the
$q_{T}$-subtraction procedure in an automatic way (see below).

Our work is organized as follows. In Sec-\ref{subsec:The-Matrix-code},
we give a short description of the MATRIX code. In Sec-\ref{subsec:Isolation-parameters},
we present the two isolation prescriptions used in the analysis. We
propose a precise estimation of the uncertainties in NNLO QCD calculations
containing at least one photon in the final state. In Sec-\ref{subsec:NNLO-Results-and},
the NNLO cross section results are presented and compared to LHC data.
We finish with the conclusion in Sec-\ref{sec:conclusion}.

\section{NNLO cross sections}

\subsection{\label{subsec:The-Matrix-code}The Matrix code}

The parton-level Monte Carlo generator MATRIX performs fully differential
computations at the next-to-next-to-leading order (NNLO) QCD, it is
based on a number of different computations and tools from various
people and groups~\cite{MATRIX-main,Matrix-1,Matrix--2,Matrix--3,Matrix--4,Matrix--5,Matrix-6-end}.
It achieves NNLO accuracy by using the $q_{T}$-subtraction formalism
in combination with the Catani\textendash Seymour dipole subtraction
method. The systematic uncertainties inherent to the $q_{T}$-subtraction
procedure may be controlled down to the few permille level or better
for all NNLO predictions. To do this, a dimensionless cut-off $r_{cut}$
is introduced which renders all cross-section pieces separately finite
and the power-suppressed contributions vanish in the limit $r_{cut}\rightarrow0$.
MATRIX simultaneously computes the cross section at several $r_{cut}$
values and then the extrapolated result is evaluated, including an
estimate of the uncertainty of the extrapolation procedure, in an
automatic way.

We can apply realistic fiducial cuts directly on the phase-space.
The core of MATRIX framework is MUNICH Monte Carlo program, allowing
to compute both QCD and EW corrections at NLO accuracy. The loop-induced
$gg$ contribution entering at the NNLO is available for the diphoton
production process.

\subsection{\label{subsec:Isolation-parameters}Isolation parameters}

An isolation requirement is necessary to prevent contamination of
the photons by hadrons produced during the collision, arising from
the decays of $\pi^{0},\eta$, etc. . Two prescriptions may be used
for this purpose: 
\begin{itemize}
\item the standard cone isolation criterion, used by collider experiments:
a photon is assumed to be isolated if, the amount of deposited hadronic
transverse energy $\sum_{_{h}}E_{T}^{^{h}}$ is smaller than some
value $E_{T}^{\textrm{max}}$, inside the cone of radius $R$ in azimutal
$\phi$ and rapidity $y$ angle centered around the photon direction:
\begin{equation}
\sum_{_{h}}E_{T}^{^{h}}\leq E_{T}^{\mathrm{max}},\quad r=\sqrt{\left(\phi-\phi_{\gamma}\right)^{2}+\left(y-y_{\gamma}\right)^{2}}\leq R.\label{eq:}
\end{equation}
\end{itemize}
$E_{T}^{\mathrm{max}}$ can be either a fixed value or a fraction
$\varepsilon$ of the transverse momentum of the photon $p_{T}^{\gamma}$:
\begin{equation}
E_{T}^{\mathrm{max}}=\mathrm{const.\qquad\mathrm{or}}\qquad E_{T}^{\mathrm{max}}=\varepsilon p_{T}^{\gamma},\quad0<\varepsilon\leq1.\label{eq:-1}
\end{equation}

$R$ and $E_{T}^{\mathrm{max}}$ are chosen by the experiment; ATLAS
and CMS use $R=0.4$, but $E_{T}^{\mathrm{max}}$ differs in their
various measurements; 
\begin{itemize}
\item the ``smooth'' cone or Frixione isolation criterion \cite{Frixione}:
in this case $E_{T}^{\mathrm{max}}$ is multiplied by a function $\chi\left(r\right)$
such that: 
\begin{equation}
\begin{cases}
\begin{array}{c}
\underset{r\rightarrow0}{\lim}\chi\left(r\right)=0\\
0<\chi\left(r\right)<1\quad\mathrm{if}\quad0<r<R
\end{array} & ;\end{cases}\label{eq:-2}
\end{equation}
\end{itemize}
a possible (and largely used) choice is 
\begin{equation}
\chi\left(r\right)=\left[\frac{1-\cos\left(r\right)}{1-\cos\left(R\right)}\right]^{n}\label{eq:-4}
\end{equation}
so that: 
\begin{equation}
\begin{cases}
\begin{array}{c}
\sum_{_{h}}E_{T}^{^{h}}\leq\chi\left[\frac{1-\cos\left(r\right)}{1-\cos\left(R\right)}\right]^{n}E_{T}^{\mathrm{max}},\\
r=\sqrt{\left(\phi-\phi_{\gamma}\right)^{2}+\left(y-y_{\gamma}\right)^{2}}\leq R,\quad\left(\mathrm{typically}\:n=1\right).
\end{array}\end{cases}\label{Frixione}
\end{equation}

Despite the fact that the Frixione criterion (formally) eliminates
all fragmentation contribution, it is not yet included in the experimental
studies. On the other hand, the use of this criterion by the theoretical
investigations at NNLO is necessary to ensure an Infra-Red (IR) safe
definition of the cross section since no fragmentation functions are
included.

In ATLAS measurement \cite{ATLAS8}, the standard criterion is adopted
for DIPHOX and \textsc{ResBos} but the ``smooth'' prescription is
used for 2\textgreek{g}NNLO, assuming $E_{T}^{\mathrm{max}}=11\:GeV$
. This is far from the Les Houches accord 2013 recommendations which
states that to match experimental conditions to theoretical calculations
with reasonable accuracy, the isolation parameters must be tight enough:$E_{T}^{\mathrm{max}}\le5\:GeV$
or $\varepsilon<0.1$ (assuming $n=1$) \cite{isolation study}.

In Ref.\cite{MATRIX-iso}, the authors presented a rather complete
study of the impact of the isolation parameters on the diphoton cross
sections. We can lift the following points from this study: 
\begin{itemize}
\item The NNLO cross sections are more sensitive to the variation of the
parameters of isolation in comparison with the NLO results, 
\item at fixed $n=1$, the total NNLO cross section for the ``smooth''
isolation increases by $6\%$ in going from $E_{T}^{\textrm{max}}=2$
to $10$ GeV, 
\item considering the interval $0.5<n<2$, at fixed $E_{T}^{\textrm{max}}=4$
GeV, the total NNLO cross section with $n=1$ increases by about $4\%$
with $n=0.5$ and decreases by about 5\% with $n=2$; the corresponding
scale uncertainty is lesser than $\pm8.7\%$. 
\end{itemize}
We notice that the isolation uncertainties due to the choice of the
isolation parameters are comparable to the scale uncertainties, thus
we have to consider the arbitrary choice of these parameters as a
major source of the theoretical systematic errors as well as uncertainties
related to the choice of the scale. This must be included in the final
result.

To evaluate these isolation uncertainties (i.e. to determine both
the central value and deviations), we use MATRIX to calculate the
NLO integrated cross sections by varying the parameters $n$ =$0.1,0.5,1,2,4,10$
and $E_{T}^{\mathrm{max}}=2,3,4,5,8,11$GeV , then the results are
compared to the NLO cross sections obtained by running the DIPHOX
code using the standard isolation prescription with the same $E_{T}^{\mathrm{max}}$
and $R$ parameters.

\begin{figure}
\raggedright{}\includegraphics[scale=0.33]{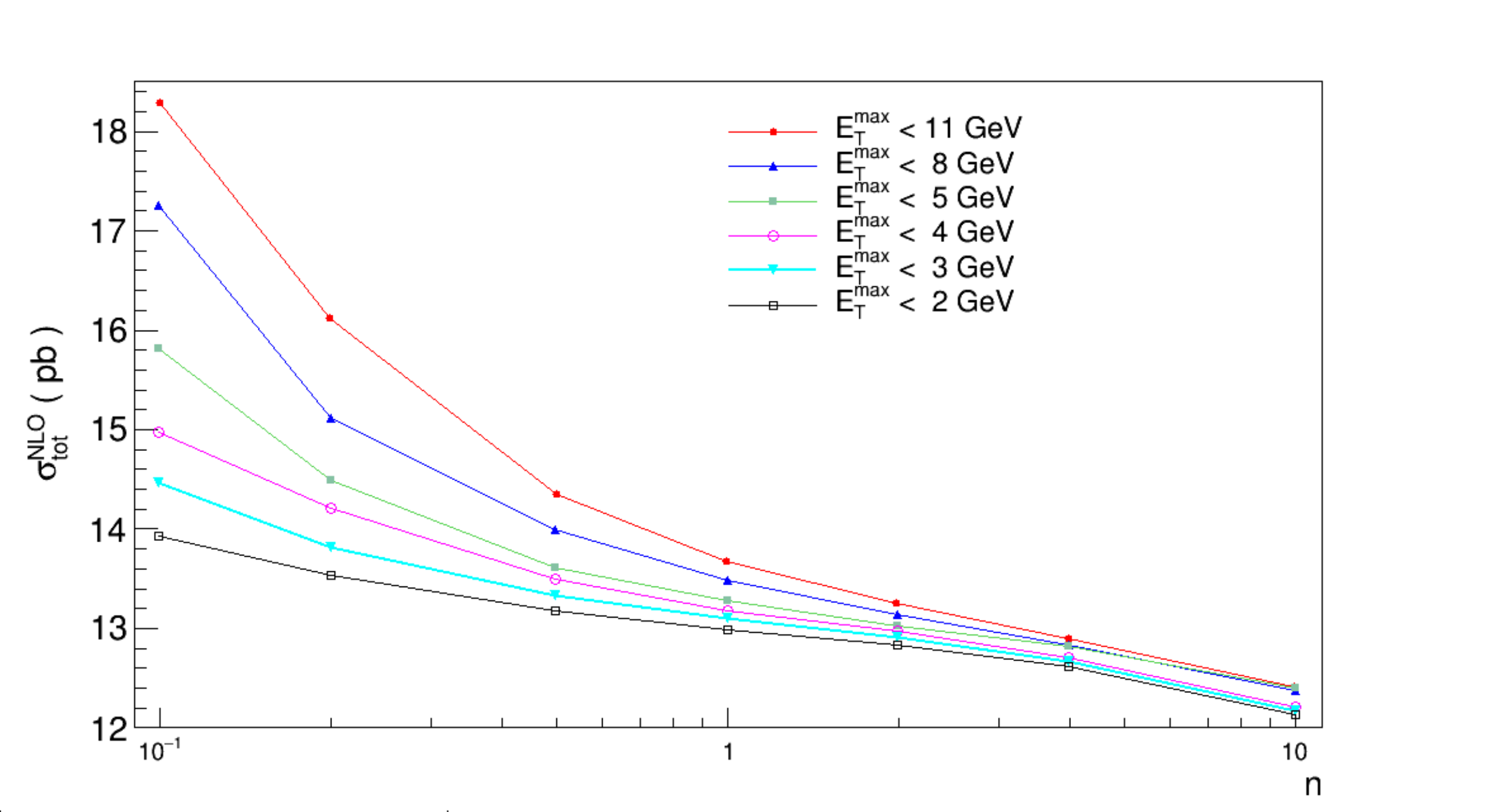}\caption{\label{fig:res-NLO_1}The MATRIX integrated fiducial cross section
$\sigma_{tot}^{NLO}$ as a function of the parameter $n$ related
to Frixione isolation criterion (see Eq.\ref{Frixione}) for different
values of $E_{T}^{\mathrm{max}}$.}
\end{figure}

\begin{figure}
\includegraphics[scale=0.41]{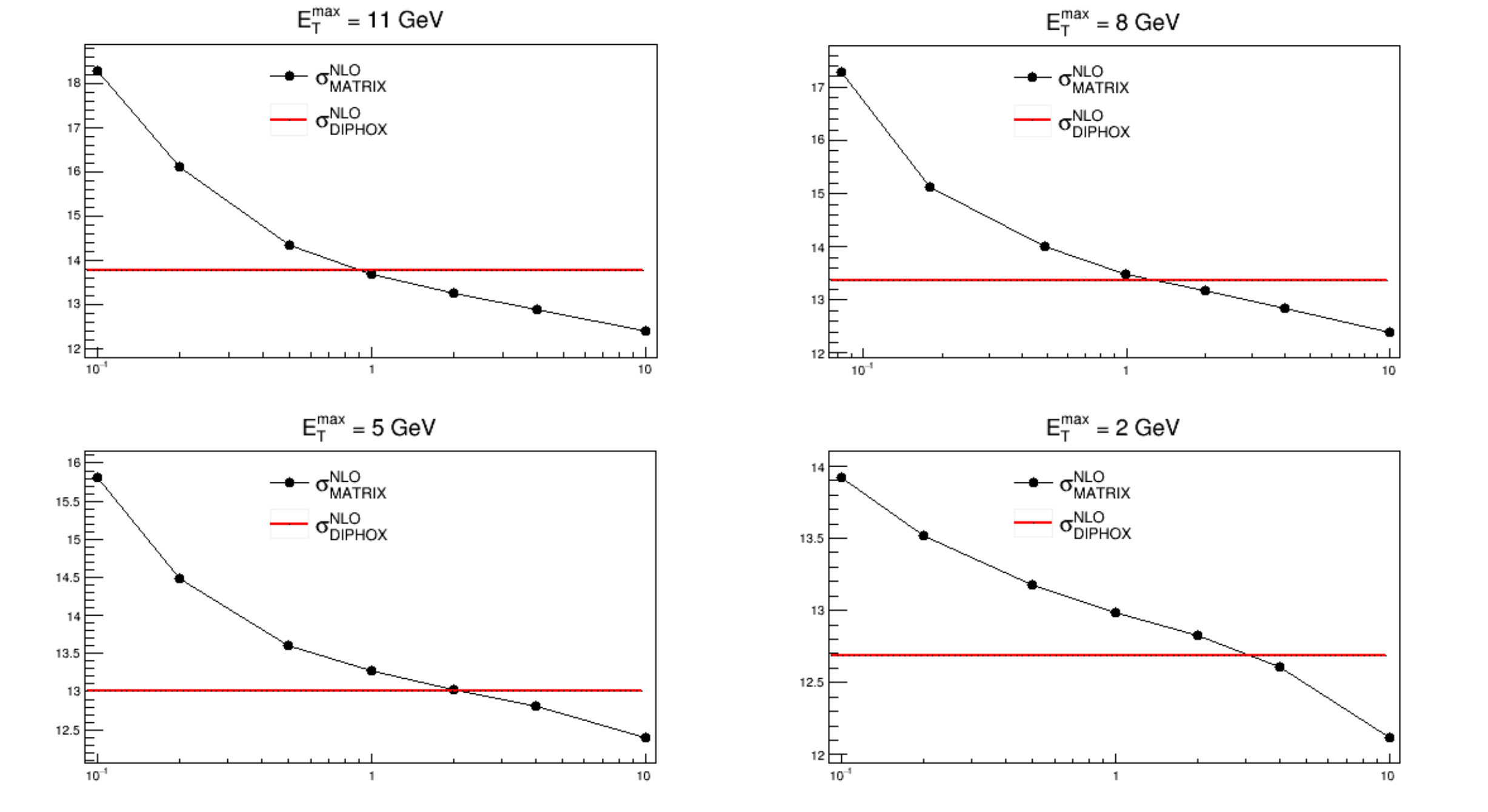}\caption{\label{fig:res-NLO_2}The MATRIX and the DIPHOX integrated fiducial
cross section $\sigma_{tot}^{NLO}$ as a function of the parameter
$n$ related to Frixione isolation criterion (see Eq.\ref{Frixione})
for several values of $E_{T}^{\mathrm{max}}$. The ``central value''
of the parameter $n=n_{0}$ depends on the value of $E_{T}^{\textrm{max}}$,
they are reported in Table \ref{tab:res-NLO}.}
\end{figure}

The so-called box (NNLO) contribution to the channel $gg\rightarrow\gamma\gamma$
is removed from the DIPHOX results to ensure that the comparaison
holds at the same NLO-order and the fine structure constant $\alpha$
is fixed to $1/137$; the setup is summarised in Table \ref{tab:setp-Diphox-Matrix}
and results are shown in Fig. \ref{fig:res-NLO_1}-\ref{fig:res-NLO_2}
.

To minimize the difference between the isolation definitions used
in the theoretical and the experimental analyses, the central value
$\sigma^{\textrm{NLO}}$ is determined at the value $n=n_{0}$ so
that: 
\begin{equation}
\sigma^{\textrm{NLO}}\equiv\left(\sigma_{\textrm{MATRIX}}^{\textrm{NLO}}\right)_{n=n_{0}}\simeq\sigma_{\textrm{DIPHOX}}^{\textrm{NLO}},\textrm{ }
\end{equation}
\[
\left(R\textrm{ and }E_{T}^{\textrm{max}}\textrm{are fixed according to the isolation experimental requirement}\right);
\]

the isolation uncertainties are evaluated by varying $n$ from $\sim\frac{1}{2}n_{0}$
to $\sim2n_{0}$. This procedure is adopted in the NNLO calculations
(see Sec-\ref{subsec:NNLO-Results-and}).

The ``central value'' of the parameter $n=n_{0}$ depends on the
value of $E_{T}^{\textrm{max}}$ (see Table \ref{tab:res-NLO}) ,
this is consistent with the results of Ref. \cite{MATRIX-iso}.

\subsection{\label{subsec:NNLO-Results-and}NNLO Results and comparaison with
data}

We consider proton\textendash proton collisions at the 8 TeV LHC.
We choose the invariant mass of the photon pair as the central scale,
i.e. 
\begin{equation}
\mu=m_{\gamma\gamma}<1700\:GeV,\label{eq:-12}
\end{equation}
Frixione isolation with $0.5<n<2,\:E_{T}^{\textrm{max}}=11\:$GeV
and $R=0.4$ (see Eq.(\ref{Frixione})), and the following fiducial
cuts: 
\begin{equation}
p_{T}^{\gamma_{1}}>40\:GeV,\quad p_{T}^{\gamma_{2}}>30\:GeV,\quad|\eta^{\gamma}|<2.37;\label{eq:-6}
\end{equation}
excluding the gap region 
\begin{equation}
1.37<|\eta^{\gamma}|<1.56.\label{eq:-7}
\end{equation}
.

The experimental angular separation between the photons is set to:
\begin{equation}
R_{\gamma\gamma}=\sqrt{\left(y_{1}-y_{2}\right)^{2}+\left(\phi_{1}-\phi_{2}\right)^{2}}>0.4,
\end{equation}
we have : 
\begin{equation}
\cosh\left(y_{1}-y_{2}\right)-\cos\sqrt{R_{\gamma\gamma}^{2}-\left(\phi_{1}-\phi_{2}\right)^{2}}\geq\left[\cosh\left(y_{1}-y_{2}\right)-\cos\sqrt{0.4^{2}-\left(\phi_{1}-\phi_{2}\right)^{2}}\right]_{\textrm{min}}\simeq0.08,
\end{equation}
and then: 
\begin{equation}
\left(m_{\gamma\gamma}\right)_{\textrm{min}}=\sqrt{2\left(p_{T}^{\gamma_{1}}\right)_{\textrm{min}}\left(p_{T}^{\gamma_{2}}\right)_{\textrm{min}}\left[\cosh\left(y_{1}-y_{2}\right)-\cos\sqrt{R_{\gamma\gamma}^{2}-\left(\phi_{1}-\phi_{2}\right)^{2}}\right]_{\textrm{min}}}\simeq13.7\textrm{ GeV}.
\end{equation}

The appropriate value of the fine structure constant $\alpha$ is
the value of the electromagnetic coupling at the invariant mass final
state $m_{\gamma\gamma}$, and since $m_{\gamma\gamma}>0$, a value
such as $\alpha_{\textrm{e.m.}}(\mu=M_{Z})$ might be more appropriate
than $\alpha_{\textrm{e.m.}}(\mu=0)\simeq1/137$. Then $\alpha$ is
fixed to $1/128.9$.

Several modern NNLO PDF sets are used (CT14 \cite{CT14 }, MMHT14
\cite{MMHT14 } and NNPDF3.1 \cite{NNPDF31}); the evolution of $\alpha_{s}$
at 3-loop order is provided by the corresponding PDF set.

For CT14, the central value of the NNLO integrated fiducial cross
section is evaluated at the isolation parameters $\left(n=n_{0}=0.84,\:E_{T}^{\textrm{max}}=11GeV\right)$
within the scale choice $\mu_{R}=\mu_{F}=m_{\gamma\gamma}$(central
scale): 
\begin{equation}
\left(\sigma_{\mathrm{tot}}^{\mathrm{fid}}\right)_{n=0.84}^{\textrm{NNLO}}=15.60\pm0.09\left(\mathrm{num}\right)\:\:\textrm{pb,}
\end{equation}
calculated at $r_{cut}$ extrapolated to zero.

The scale uncertainties are estimated in the usual way by independently
varying $\mu_{R}$ and $\mu_{F}$ in the range 
\begin{equation}
\frac{1}{2}m_{\gamma\gamma}\le\mu_{R},\:\mu_{F}\le2m_{\gamma\gamma},\label{eq:-8}
\end{equation}
with the constraint 
\begin{equation}
\frac{1}{2}\le\mu_{R}/\mu_{F}\le2.\label{eq:-9}
\end{equation}

The relative scale uncertainty in the integrated cross section is
$\left(_{-5.6\%}^{+6.7\%}\right)$.

The relative isolation uncertainty (at the central scale) is calculated
by varying $n$ from $0.5$ to $2$: 
\begin{equation}
\begin{cases}
\begin{array}{c}
\frac{\sigma_{n=0.5}-\sigma_{n=0.84}}{\sigma_{n=0.84}}\simeq+3.8\%\\
\frac{\sigma_{n=2}-\sigma_{n=0.84}}{\sigma_{n=0.84}}\simeq-5.5\%
\end{array}\end{cases}
\end{equation}
The impact of the variation of the strong coupling constant is also
investigated. The change of $\alpha_{s}\left(M_{Z}^{2}\right)$ by
$\pm0.001$ from the central value $0.118$ leads to variations $\left(_{-1.0\%}^{+0.6\%}\right)$
in the fiducial integrated cross section. The cross sections related
to CT14, MMHT14 and NNPDF3.1 modern PDF sets are very close to each
other with an uncertainty less than $0.4\%$.

We can write our theoretical prediction of the integrated fiducial
cross section as:

\begin{align}
\sigma_{\mathrm{tot}}^{\mathrm{fid}}\simeq & 15.60\pm0.09\left(\mathrm{num}\right)\:{}_{-5.7\%}^{+6.7\%}\:\mathrm{\left(scale\right)\:{}_{-5.5\%}^{+3.8\%}\:\mathrm{\left(iso\right)}}\\
\simeq & 15.60\pm0.09\left(\mathrm{num}\right)\:{}_{-0.89}^{+1.05}\:\mathrm{\left(scale\right)\:{}_{-0.86}^{+0.59}\:\mathrm{\left(iso\right)}}\nonumber \\
\simeq & 15.60_{-1.24}^{+1.21}\simeq\left(15.6\pm1.2\right)\textrm{pb}\nonumber 
\end{align}
which is consistent with the experimental data \cite{ATLAS8}:$\left(16.8\pm0.8\right)\textrm{pb.}$

Note that the theoretical uncertainties are dominated by both the
scale and the isolation systematic errors which are of the same order.

Since this process involves isolated photons in the final state it
has a relatively large numerical uncertainty at NNLO after the $r_{cut}\rightarrow0$
extrapolation, and as recommended by authors of Ref.\cite{MATRIX-main},
the distribution calculated at fixed $r_{cut}=0.05\%$ must be multiplied
by the correction factor: 
\begin{equation}
\frac{\left(\sigma_{\mathrm{tot}}^{\mathrm{fid}}\right)_{r_{cut}\rightarrow0}}{\left(\sigma_{\mathrm{tot}}^{\mathrm{fid}}\right)_{r_{cut}=0.05\%}}\left(\sim0.98\right).\label{eq:-3}
\end{equation}

\begin{figure}
\centering{}\includegraphics[scale=0.43]{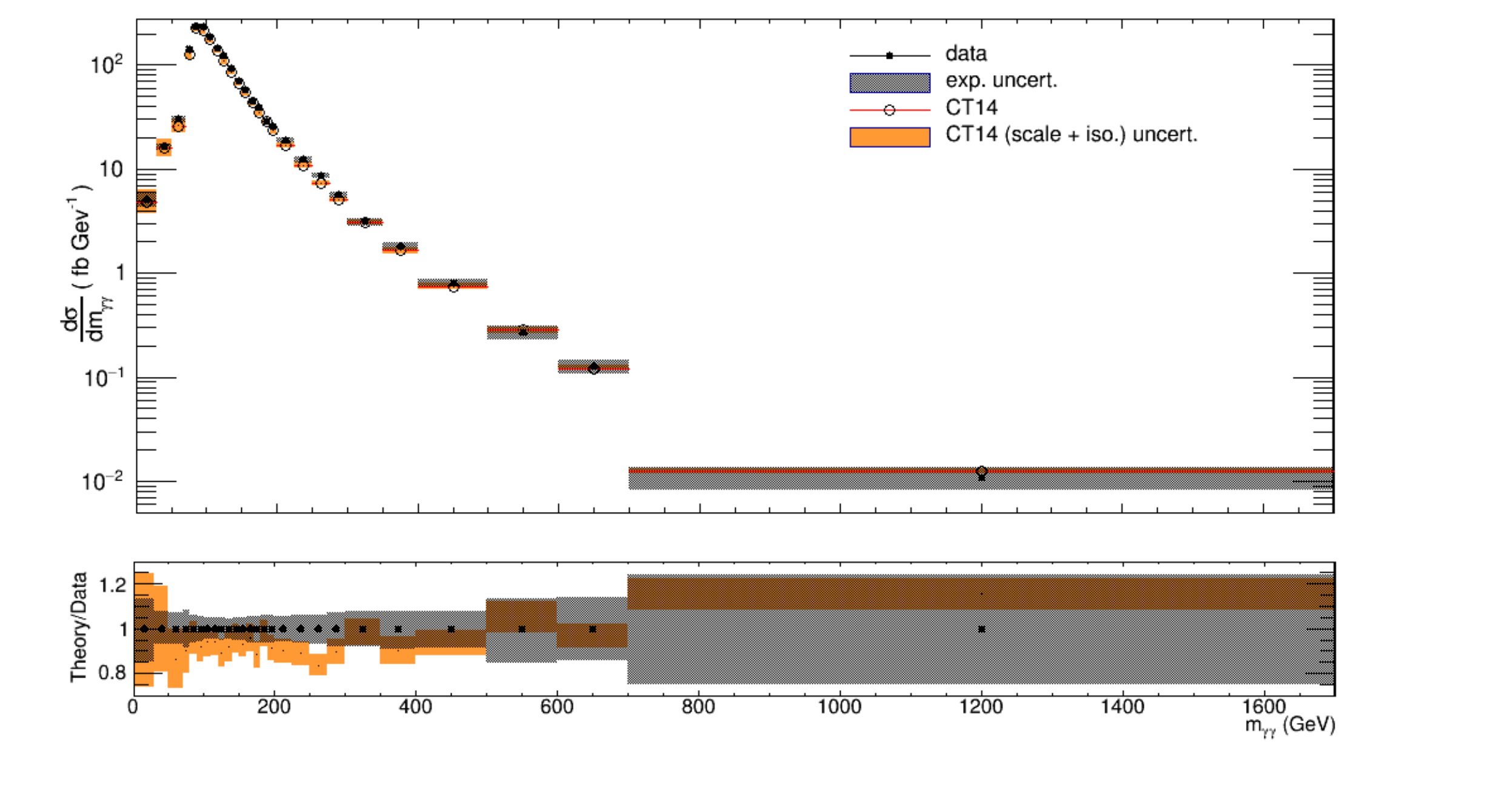}\caption{\label{fig:all-1700}The MATRIX differential fiducial cross section
related to CT14 as a function of $m_{\gamma\gamma}$ compared to the
data \cite{ATLAS8}. }
\end{figure}

\begin{figure}
\centering{}\includegraphics[scale=0.43]{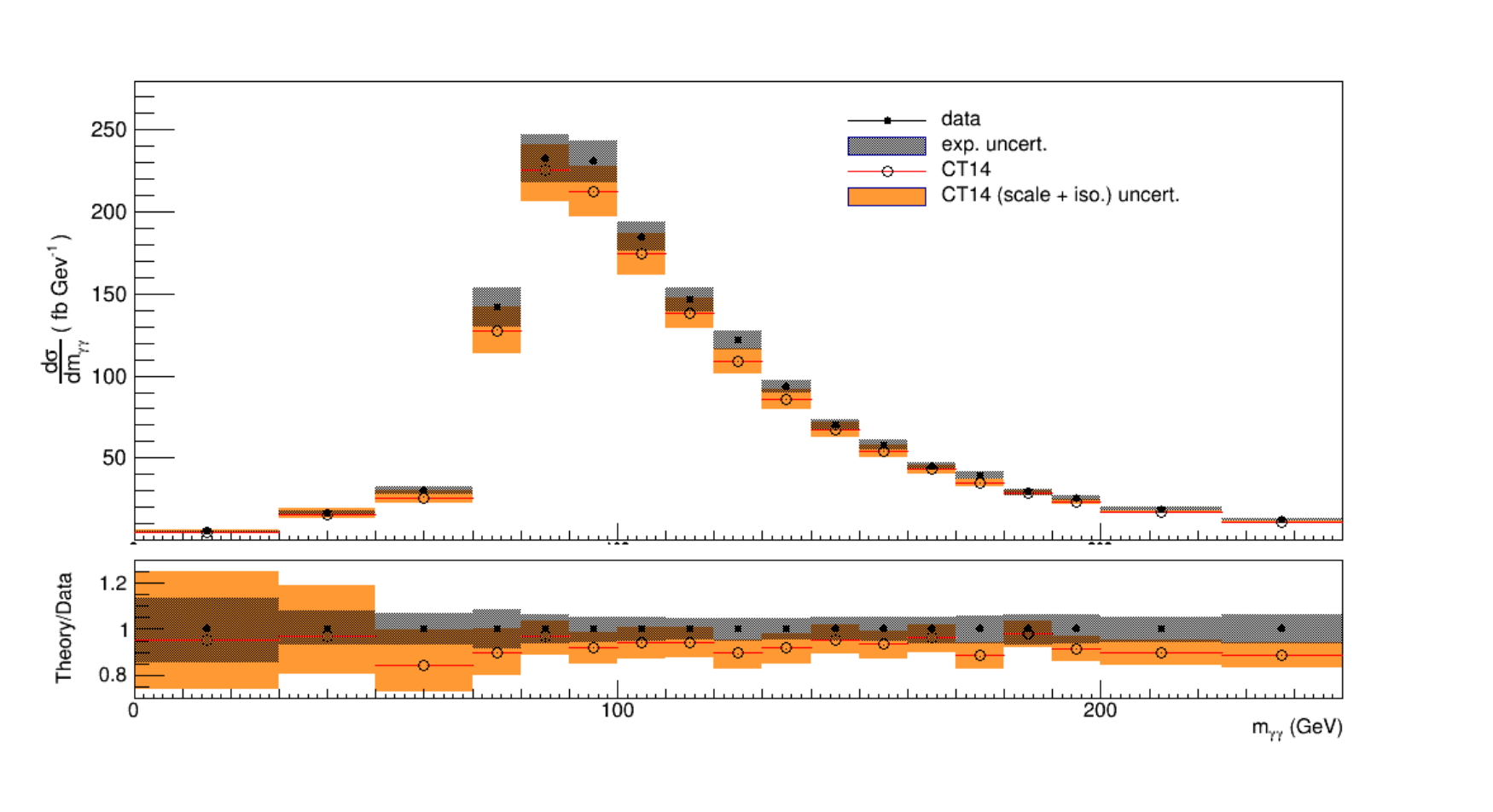}\caption{\label{fig:all-250}The MATRIX differential fiducial cross section
related to CT14 as a function of $m_{\gamma\gamma}$ compared to the
data \cite{ATLAS8}, in the range $0<m_{\gamma\gamma}<250\:GeV.$}
\end{figure}

The MATRIX differential cross section is consistent with data as shown
in Fig.\ref{fig:all-1700}-\ref{fig:all-250}.

\section{\label{sec:conclusion}Conclusion}

We presented the calculation of the integrated and differential cross
sections for the isolated diphoton production in pp collisions at
the centre\textendash of\textendash mass energy $\sqrt{s}=8$ TeV
in next-to-next-to-leading order (NNLO) QCD using the computational
framework MATRIX. A special care was paid to the choice of the Frixione
isolation parameters. We kept the same value of $E_{T}^{\mathrm{max}}=11\:GeV$
and $R=0.4$ used by experimentalists but we adjusted the value of
the parameter $n$ until the integrated cross section calculated by
MATRIX matches that calculated by DIPHOX at the same NLO-order ( without
the \textit{Box}-contribution to the channel $gg\rightarrow\gamma\gamma$).

Once these parameters were fixed, we calculated the central value
of the MATRIX (NNLO) cross sections and by varying the Frixione parameter
$n$ from 0.5 to 2, we estimated the relative isolation uncertainty
$\left(_{-5.5\%}^{+3.8\%}\right)$. The scale uncertainty is found
to be equal to$\left(_{-5.7\%}^{+6.7\%}\right)$.

Both the scale and the isolation uncertainties were of the same order
and represent the main source of the theoretical errors, the uncertainties
inherent to the $q_{T}$-subtraction procedure $\left(\sim0.6\%\right)$
and to the variation of the coupling constant $\alpha_{s}\left(M_{Z}^{2}\right)$
$\left(\sim0.8\%\right)$ were negligible.

Our predictions for the differential and the integrated cross sections
are in good agreement with the data. In particular we have 
\[
\sigma_{\mathrm{tot}}^{\mathrm{fid}}\simeq15.60\pm0.09\left(\mathrm{num}\right)\:{}_{-5.7\%}^{+6.7\%}\:\mathrm{\left(scale\right)\:{}_{-5.5\%}^{+3.8\%}\:\mathrm{\left(iso\right)}}\simeq\left(15.6\pm1.2\right)\textrm{pb}.
\]

\section*{Acknowledgements}

This work was realized with the support of the FNR (Algerian Ministry
of Higher Education and Scientic Research), as part of the research
project D018 2014 0044. We gratefully acknowledge computing support
provided by the Research Center on Scientific and Technical Information
(CERIST) in Algiers (Algeria) through the HPC platform \textsc{ibnbadis}. 

\begin{table}
\centering{}%
\begin{tabular}{|c||c|}
\hline 
DIPHOX v.1.2  & MATRIX v.1.0\tabularnewline
\hline 
pdf\cite{CTEQ6}:cteq6  & cteq6\tabularnewline
$\alpha$ fixed to $1/137$  & $\alpha$ fixed to $1/137$\tabularnewline
$p_{T}^{\gamma}>25\:GeV,\quad|\eta^{\gamma}|<2.37;$  & $p_{T}^{\gamma}>25\:GeV,\quad|\eta^{\gamma}|<2.37;$\tabularnewline
$80<m_{\gamma\gamma}<1700$ GeV  & $80<m_{\gamma\gamma}<1700$ GeV\tabularnewline
isolation: $R=0.4$, standard, $E_{T}^{\textrm{max}}$.  & $R=0.4$, ``smooth'', $(E_{T}^{\textrm{max}},n)$\tabularnewline
fragmentation functions\cite{BFG-II}: BFG set II  & -\tabularnewline
The direct part: born only,no box contributions  & -\tabularnewline
\hline 
\end{tabular}\caption{\label{tab:setp-Diphox-Matrix}Setup of the diphoton production process
used in the NLO runs.}
\end{table}

\begin{table}
\centering{}%
\begin{tabular}{|c|c|c|}
\hline 
$E_{T}^{\textrm{max}}$(GeV)  & $n_{0}$  & $\sigma_{MATRIX}^{NLO}$(pb)\tabularnewline
\hline 
\hline 
11  & 0.84  & $13.78\pm0.12\textrm{(num}\textrm{)}_{-5.0\%}^{+6.1\%}\textrm{(scale})$\tabularnewline
\hline 
8  & 1.2  & $13.36\pm0.10\textrm{(num}\textrm{)}_{-4.8\%}^{+5.9\%}\textrm{(scale})$\tabularnewline
\hline 
5  & 2.0  & $13.01\pm0.10\textrm{(num}\textrm{)}_{-4.7\%}^{+5.8\%}\textrm{(scale})$\tabularnewline
\hline 
2  & 3.2  & $13.69\pm0.11\textrm{(num}\textrm{)}_{-4.6\%}^{+5.7\%}\textrm{(scale})$\tabularnewline
\hline 
\end{tabular}\caption{\label{tab:res-NLO}The ``central value'' of the parameter $n=n_{0}$.}
\end{table}

\begin{thebibliography}{10}
\bibitem{DIPHOX} T. Binoth, J. P. Guillet, E. Pilon and M. Werlen,
Eur. Phys. J. C16, 311 (2000).

\bibitem{ResBos} C. Balazs, E. L. Berger, P. M. Nadolsky and C. -P.
Yuan, Phys. Rev. D76, 013009 (2007).

\bibitem{2=00003D00003D0003B3Res}L. Cieri, F. Coradeschi, and D.
de Florian, JHEP 06 (2015) 185, {[}arXiv :1505.0316{]}.

\bibitem{2=00003D00003D0003B3NNLO} S. Catani,L. Cieri, D. de Florian,
G. Ferrera and M. Grazzini, Phys. Rev. Lett. 108 (2012) 072001,{[}arXiv:1110.2375{]}.

\bibitem{MCFM} R. Boughezal, J. M. Campbell, R. K. Ellis, C. Focke,
W. Giele, X. Liu, F. Petriello, and C. Williams, Color singlet production
at NNLO in MCFM, Eur. Phys. J. C77 (2017), no. 1 7, {[}arXiv:1605.08011{]}.

\bibitem{MATRIX-main}M. Grazzini, S. Kallweit and M. Wiesemann, arXiv:1711.06631
{[}hep-ph{]}. Matrix is available for download from: http://matrix.hepforge.org/

\bibitem{ATLAS7}ATLAS Collaboration, JHEP 01 (2013) 086, arXiv: 1211.1913
{[}hep-ex{]}.

\bibitem{CMS7}CMS Collaboration, Eur. Phys. J. C 74 (2014) 3129,
arXiv:1405.7225 {[}hep-ex{]}.

\bibitem{ATLAS8}ATLAS Collaboration, Phys. Rev. D 95, 112005 (2017),
arXiv:1704.03839 {[}hep-ex{]}.

\bibitem{Matrix-1}C. Anastasiou, E. W. Nigel Glover, and M. E. Tejeda-Yeomans.
Two loop QED and QCD corrections to massless fermion boson scattering.
Nucl. Phys., B629:255289, 2002.

\bibitem{Matrix--2}Fabio Cascioli and Maierh.

\bibitem{Matrix--3}Stefano Catani, Leandro Cieri, Daniel de Florian,
Giancarlo Ferrera, and Massimiliano Grazzini. Diphoton production
at hadron colliders: a fully-dierential QCD calculation at NNLO. Phys.
Rev. Lett., 108:072001, 2012. {[}Erratum: Phys. Rev. Lett.117,no.8,089901(2016){]}.

\bibitem{Matrix--4}Stefano Catani, Leandro Cieri, Daniel de Florian,
Giancarlo Ferrera, and Massimiliano Grazzini. Vector boson production
at hadron colliders: hard-collinear coecients at the NNLO. Eur. Phys.
J., C72:2195, 2012.

\bibitem{Matrix--5}Stefano Catani and Massimiliano Grazzini. An NNLO
subtraction formalism in hadron collisions and its application to
Higgs boson production at the LHC. Phys. Rev. Lett., 98:222002, 2007.

\bibitem{Matrix-6-end}Ansgar Denner, Stefan Dittmaier, and Lars Hofer.
Collier: a fortran-based Complex One-Loop LIbrary in Extended Regularizations.
Comput. Phys. Commun., 212:220238, 2017.

\bibitem{Frixione}S. Frixione, Phys. Lett. B429 (1998) 369, {[}hep-ph/9801442{]}.

\bibitem{isolation study}Leandro Cieria, arXiv:1510.06873 {[}hep-ph{]}.

\bibitem{MATRIX-iso}S.Catani, L.Cieri, D.de Florian, G.Ferrera and
M.Grazzini, arXiv:1802.02095 {[}hep-ph{]}.

\bibitem{CT14 }S. Dulat et al., Phys. Rev. D 93, no. 3, 033006 (2016),
arXiv:1506.07443 {[}hep-ph{]}.

\bibitem{MMHT14 }L. A. Harland-Lang, A. D. Martin, P. Motylinski
and R. S. Thorne, Eur. Phys. J. C 75, no. 5, 204 (2015), arXiv:1412.3989
{[}hep-ph{]}.

\bibitem{NNPDF31}The NNPDF Collaboration, arXiv:1706.00428 {[}hep-ph{]}.

\bibitem{CTEQ6}J. Pumplin, D.R. Stump, J. Huston, H.L. Lai, P. Nadolsky
and W.K. Tung, J HEP 0207:012,2002,arXiv:hep-ph/0201195.

\bibitem{BFG-II}L. Bourhis, M. Fontannaz and J. P. Guillet, Eur.
Phys. J. C2 (1998) 529\textendash 537, arXiv:hep-ph/9704447. \pagebreak{}
\end{thebibliography}
\end{document}